\pdfoutput=1
\documentclass[12pt]{article}

\usepackage{graphicx}
\usepackage{xcolor}
\usepackage[unicode,breaklinks=true,linktocpage=true,pagebackref=true]{hyperref}

\usepackage{ascmac}

\usepackage{amsmath,amssymb,amsthm}
\usepackage{amscd} 
\usepackage[all]{xy}
\usepackage{mathrsfs} 
\numberwithin{equation}{section}

\usepackage[top=0.14\paperwidth,bottom=0.14\paperwidth,left=0.14\paperwidth,right=0.14\paperwidth]{geometry}

\usepackage{cite}

\renewcommand*{\backref}[1]{}
\renewcommand*{\backrefalt}[4]{%
  \ifcase #1 %
{\bf \color{red}No citations.} 
  \or
    (page #2).%
  \else
    (pages #2).%
  \fi%
}

\DeclareMathOperator{\link}{Link}

\newcommand{\ig}{\includegraphics}

\newcommand{\vevs}[1]{\langle #1 \rangle}

\newcommand{\er}[1]{Eq.~\eqref{#1}}
\newcommand{\ers}[1]{Eqs.~\eqref{#1}}
\newcommand{\qtq}[1]{\quad\text{#1}\quad}

\newcommand{\bb}{\mathbb}
\newcommand{\na}{\nabla}

\renewcommand{\b}{\bar}

\newcommand{\pd}[2]{\frac{\partial{#1}}{\partial{#2}}}

\newcommand{\bs}{\boldsymbol}

\newcommand{\fr}{\frac}

\newcommand{\der}{\partial}

\renewcommand{\(}{\left(}
\renewcommand{\)}{\right)}

\newcommand{\wed}{\wedge}
\newcommand{\weds}{\wedge \cdots \wedge}
\newcommand{\bmx}{\left(\begin{matrix}}
\newcommand{\emx}{\end{matrix}\right)}

\begin{document}

\allowdisplaybreaks[4]
\begin{titlepage}
\hfill KEK-TH-2481

\renewcommand{\thefootnote}{\fnsymbol{footnote}}%
 \vspace{3em}
 \begin{center}
  {\Large
Non-invertible symmetries 
in axion electrodynamics
  \par
   }
 \vspace{1.5em}
  {\large
    Ryo Yokokura\footnote{ryokokur@post.kek.jp}${}^{a,b}$
   \par
   }
  \vspace{1em} 
${}^a${\small\it KEK Theory Center, Tsukuba 305-0801, Japan}
\\
${}^b${\small\it Research and Education Center for Natural Sciences,
\par 
Keio University, Hiyoshi 4-1-1, Yokohama, Kanagawa 223-8521, Japan}
\par
  \vspace{1em} 
 \end{center}
 \par
\vspace{1.5em}
\begin{abstract}
We study non-invertible global symmetries in $(3+1)$-dimensional 
axion electrodynamics with a massless axion and a massless photon.
In addition to a previously known 
non-invertible 0-form shift symmetry of the axion, 
we find a non-invertible 1-form symmetry associated with 
the equation of motion for the photon.
Correlation functions of non-invertible symmetry defects
lead to invertible 1- and 2-form symmetry defects 
associated with Bianchi identities 
for the axion and photon.
In terms of the correlation functions, we discuss 
several phenomena for extended objects, such as
induced fractional electric charges on axionic domain walls 
and fractional axionic operators on intersection points 
of magnetic flux tubes from the viewpoint of 
global symmetries.
\end{abstract}
\end{titlepage}
\setcounter{footnote}{0}
\renewcommand{\thefootnote}{$*$\arabic{footnote}}

\tableofcontents

\section{Introduction}
Axions have been extensively 
studied in various fields of modern physics.
Examples include
a candidate to resolve the strong CP problem in particle 
physics~\cite{Peccei:1977hh,Weinberg:1977ma,Wilczek:1977pj,Dine:1981rt,Zhitnitsky:1980tq,Kim:1979if,Shifman:1979if},
a candidate of cold dark matter~\cite{Preskill:1982cy,Abbott:1982af,Dine:1982ah,Stecker:1982ws,Masso:1995tw}, an inflaton in cosmology~\cite{Freese:1990rb,Adams:1992bn},
moduli fields in low-energy effective theories 
of string theory~\cite{Witten:1985xb,Townsend:1993wy,Izquierdo:1993st,Harvey:2000yg,Svrcek:2006yi,Arvanitaki:2009fg},
and 
quasi-particle excitations in topological materials~\cite{Wilczek:1987mv,Qi:2008ew,Essin:2008rq,Li:2009tca,Wang:2012bgb}.
One of the characteristic features of the axions is 
a topological coupling to the photon.
This coupling arises from the chiral anomaly
of massive fermions in ultraviolet theories
coupled with the axion and photon.
The axion-Maxwell theory with the topological coupling is called 
the axion electrodynamics~\cite{Wilczek:1987mv}.

It has been known that 
the axion electrodynamics exhibits peculiar phenomena 
for extended objects, including 
axionic domain walls and magnetic flux tubes,
because the equations of motion are modified
by the axion-photon coupling.
The electric Gauss law 
with the axion-photon coupling leads to an effect that we will call 
Sikivie effect: If an axionic domain wall 
intersects with a magnetic flux tube, 
there is an induced electric charge in the intersection
\cite{Sikivie:1984yz}. 
The modification of Maxwell-Amp\`ere law 
gives rise to the anomalous Hall effect on the axionic domain wall:
When we add an electric flux on the axionic domain wall, 
we have induced electric current.
The direction of the current is perpendicular to both electric flux and the normal vector of the axionic domain wall~\cite{Sikivie:1984yz, Wilczek:1987mv, Qi:2008ew, Teo:2010zb}.
The equation of motion for the axion implies that the axion can be 
created by the intersection of two moving magnetic flux tubes~\cite{Qi:2012cs},
because moving magnetic tubes also have 
electric flux by the Lorentz boost.
One may ask a question whether there is any underlying structure 
in the above phenomena on extended objects.

The notion of higher-form global symmetries 
is one of the candidates for understanding the structure.
Here, higher $p$-form symmetries are symmetries 
whose charged objects are $p$-dimensional extended objects~\cite{Banks:2010zn, Kapustin:2014gua, Gaiotto:2014kfa} 
(see also Refs.~\cite{Batista:2004sc,Pantev:2005zs,Pantev:2005wj, Pantev:2005rh, Nussinov:2006iva, Nussinov:2008aa, Nussinov:2009zz, Nussinov:2011mz, Distler:2010zg} for earlier works and Refs.~\cite{Kapustin:2013uxa,Kapustin:2013qsa,Gaiotto:2017yup,Gaiotto:2017tne,Tanizaki:2017qhf,Tanizaki:2017mtm,Komargodski:2017dmc, deAlmeida:2017dhy,Benini:2018reh,Cordova:2018cvg, Delcamp:2018wlb, Wen:2018zux,Hirono:2018fjr,Delcamp:2019fdp, Hirono:2019oup,Hidaka:2019jtv,Anber:2019nze,Misumi:2019dwq,Hidaka:2019mfm,Anber:2020xfk,Anber:2020gig,Hidaka:2020ucc,Yamamoto:2020vlk,Furusawa:2020kro,Hsin:2020nts,Gukov:2020btk,Iqbal:2020lrt,Brauner:2020rtz,DeWolfe:2020uzb,Brennan:2020ehu,Heidenreich:2020pkc,Apruzzi:2021vcu,Bhardwaj:2021wif,Yamamoto:2022vrh,Barkeshli:2022wuz,Barkeshli:2022edm} for recent progress).
Symmetry generators of $p$-form symmetries can be identified as 
codimension-$(p+1)$ objects, which are called symmetry defects.
Ordinary symmetries can be interpreted as 0-form symmetries
because they act on local 0-dimensional operators,
and conserved charges are codimension-1 objects.
One of the important properties of symmetry defects is
that they are topological, i.e., they can move freely.
This property is a generalization of the conservation laws of 
ordinary symmetry generators.

In the case of the axion electrodynamics, 
it has been shown that there are 0-, 1-, and 2-form 
global symmetries in the case of the massless axion and photon~\cite{Hidaka:2020iaz,Hidaka:2020izy}. 
It has been found that the global symmetries are extended to
0-, 1-, 2-, and 3-form symmetries in the case of the 
low-energy limit of the massive axion
and photon system~\cite{Hidaka:2021mml,Hidaka:2021kkf}
(see also higher-form symmetries in axion-Yang-Mills theories~\cite{Brennan:2020ehu} and in higher-dimensional axion electrodynamics~\cite{Nakajima:2022feg}).
The symmetries can be classified as electric symmetries and magnetic symmetries~\cite{Gaiotto:2014kfa}.
On the one hand, 
electric symmetries are associated with the equations of motion 
for the axion and photon, which include the electric Gauss law.
On the other hand, 
the magnetic symmetries are obtained by the Bianchi identities,
which include the magnetic Gauss law.
The electric 0-form symmetry defect can be identified as a worldvolume of an axionic domain wall 
as a codimension-1 defect.
Similarly, we can understand the electric 1-form symmetry defect  as 
an integral of an electric flux on a temporally localized 
codimension-2 defect or a magnetic flux on a temporally extended codimension-2 defect.
It has been found that 
we can understand 
the Sikivie effect, anomalous Hall effect, 
and the axion production on magnetic fluxes 
from the viewpoint of the correlation functions of 
symmetry defects.

However, there are some remaining issues in the above discussion by
the higher-form symmetries.
Due to the chiral anomaly,
we can only take finite values for the domain walls and magnetic flux
as the 0- and 1-form electric symmetry defects, respectively.
In particular, 
the groups of the electric 0- and 1-form symmetries should be $\bb{Z}_N$
in the case of the massless axion and photon.
Here, the integer $N$ is the coefficient in 
the axion-photon coupling, which is roughly the number of 
massive Dirac fermions that are coupled to the axion and photon in 
ultraviolet theories.
It is not straightforward to consider, e.g., 
a magnetic flux with a smaller value than $2\pi /N$
from in the language of the higher-form symmetries.

Recently, 
it has been shown that the chiral symmetry that is violated
by the chiral anomaly in Abelian gauge theories
can be modified so that it is still a symmetry
for any rotational angle $e^{2\pi i r}$ with a rational number 
$r$~\cite{Choi:2022jqy,Cordova:2022ieu}.
We can modify the gauge non-invariant topological term 
in the current of the chiral symmetry
using a topological quantum field theory (TQFT) describing 
a fractional quantum Hall state~\cite{Wen:1989zg,Wen:1989iv}.
Due to the presence of the TQFT, 
the symmetry defects become non-invertible.
Such symmetries are called non-invertible symmetries, 
which are extensively studied in recent years~\cite{Bhardwaj:2017xup,Tachikawa:2017gyf,Nguyen:2021yld,Nguyen:2021naa,Heidenreich:2021xpr,Sharpe:2021srf,Koide:2021zxj,Inamura:2021szw,Kaidi:2021xfk,Choi:2021kmx,Cordova:2022rer,Roumpedakis:2022aik,Hayashi:2022fkw,Choi:2022zal,Kaidi:2022uux,Antinucci:2022eat,Inamura:2022lun,Damia:2022rxw,Damia:2022bcd,Choi:2022rfe,Lin:2022xod,Niro:2022ctq,Antinucci:2022vyk,Kaidi:2022cpf,Chen:2022cyw,Cordova:2022fhg,Karasik:2022kkq,GarciaEtxebarria:2022jky}.
One of the virtues of this non-invertible symmetry is 
that we can take an arbitrarily small rational number as 
a rotational angle.
By applying the non-invertible symmetry to axion electrodynamics
with the massless axion and photon, 
the non-invertible 0-form symmetry has been obtained~\cite{Cordova:2022ieu}. 
The corresponding symmetry defect can describe a worldvolume of
axionic domain wall, which can generate a shift 
transformation on the axion with an arbitrarily small rational number.
Therefore, one can expect that the non-invertible symmetries 
can resolve the above-mentioned issues if we can construct 
similar non-invertible symmetry for the electric 1-form symmetry.

In this paper, we investigate non-invertible symmetries in 
$(3+1)$-dimensional axion electrodynamics.
We construct a non-invertible electric 1-form symmetry defect
with a rational parameter. 
This can be done by modifying the topological term 
of the invertible 1-form symmetry generator using 
a TQFT that describes 
a low-energy effective theory of $(1+1)$-dimensional 
Schwinger model~\cite{Coleman:1974bu,Anber:2018jdf,Armoni:2018bga,Misumi:2019dwq}.
In contrast to the invertible 1-form symmetry parameterized 
by a finite group $\bb{Z}_N$, our non-invertible 1-form symmetry 
can be parameterized by a rational number which can be 
arbitrarily small.

We then discuss the correlation functions of non-invertible 
the 0- and 1-form symmetry defects.
We show that an intersection of the 
0- and 1-form symmetry defects leads to 
a magnetic 1-form symmetry defect
which have a boundary at the intersection.
The intersection of the defects is transformed under the action of the non-invertible 1-form symmetry.
The transformation shows that there is an induced electric charge or
current on the intersections captured by the electric Gauss law
and Maxwell-Amp\`ere law, 
which means the Sikivie effect and the anomalous Hall effect.
Similarly, an intersection of two non-invertible 
1-form symmetry defects 
induces a magnetic 2-form symmetry defect
which has a boundary at the intersection.
We show that the intersection is transformed under the 
action of the non-invertible 
0-form symmetry.
The transformation implies that there 
is an axion on the intersection,
which 
can be understood as the axion production on the magnetic flux.

This paper is organized as follows.
In section~\ref{inv}, we review invertible higher-form symmetries 
in the axion electrodynamics.
We then review the non-invertible 0-form symmetry in section~\ref{ni0}.
In section~\ref{ni1}, we propose the non-invertible 1-form symmetry.
By using the non-invertible symmetries, 
we investigate the correlation functions of symmetry defects, 
and 
discuss their relations to the Sikivie effect, anomalous Hall effect, and 
the production of the axion on the magnetic flux
in section~\ref{corr}.
Finally, we summarize this paper in section~\ref{sum}.
Continuous deformations and 
symmetry transformation of non-invertible defects
are discussed in appendix~\ref{top}.

{\it Note added}: 
While this work was being completed,
we received Ref.~\cite{2212.04499}, 
where the non-invertible 1-form symmetry 
in the axion electrodynamics was considered.
There are some overlaps in 
the construction of non-invertible 1-form symmetry defects 
as well as the correlation functions of two non-invertible 
symmetry defects between this paper and Ref.~\cite{2212.04499}.%
\footnote{The reference~\cite{2212.04499} contains excellent
results, e.g., the selection rules for the monopoles and axionic strings
based on non-invertible symmetries.}
Meanwhile, 
we apply the non-invertible symmetries 
to the phenomena for extended objects e.g., the Sikivie effect by using 
correlation functions of three non-invertible 
symmetry defects (see section~\ref{corr3} and \ref{phys}).

\section{Invertible symmetries in axion electrodynamics\label{inv}}

In this section, we review invertible higher-form global symmetries
in the axion electrodynamics~\cite{Hidaka:2020iaz,Hidaka:2020izy}.
Two of them are associated with the equations of motion 
for the axion and photon, and the rest of them are derived by 
the Bianchi identities.

In this paper, we use the notation of differential forms: 
$d$ is the exterior differential, $\wed$ denotes the wedge product, 
$*$ represents the Hodge star operator,
$\int |\omega_p|^2 = \int \omega_p \wed *\omega_p
 = \int d^4 x \omega^{\mu_1...\mu_p}\omega_{\mu_1...\mu_p} $
for a $p$-form field $\omega_p$.
We use the spacetime metric $\eta_{\mu\nu} = (-1,+1,+1,+1)$
and the totally anti-symmetric tensor $\epsilon_{\mu\nu\rho\sigma}$
with $\epsilon_{0123} = +1$.
\subsection{Action}

We consider the axion electrodynamics with a massless axion and 
a massless photon.
We introduce the following action~\cite{Wilczek:1987mv},
\begin{equation}
 S [\phi, a]
= 
- \int 
\(
\fr{v^2}{2} |d\phi|^2 + \fr{1}{2e^2} |da|^2 - \fr{N}{8\pi^2} \phi da \wed da
\).
\end{equation}
Here, $\phi$ is the axion, $a$ is the photon,
$v$ is a decay constant, $e$ is a coupling constant,
and $N$ is an integer.
We assume that the spacetime is a spin manifold.
The axion is assumed to be a $2\pi$ periodic 0-form 
$\phi ({\cal P})+ 2\pi \sim \phi ({\cal P})$ 
at a point ${\cal P}$ in the spacetime.
By this periodicity, the line integral of the axion 
should be quantized as $\int_{\cal C} d\phi \in 2\pi \bb{Z}$
for a closed 1-dimensional subspace ${\cal C}$.
The quantization physically means that there can exist
a worldsheet of an axionic string.
The photon $a = a_\mu dx^\mu$ 
is a $U(1)$ 1-form gauge field 
with a gauge transformation $a \to a + d\lambda$
by a gauge parameter 0-form $\lambda$.
Here, the gauge field and its gauge parameter are quantized 
as $\int_{\cal S} da \in 2\pi \bb{Z}$ and 
$\int_{\cal C} d\lambda \in 2\pi \bb{Z}$
on a closed 2-dimensional subspace ${\cal S}$.
The quantization of the photon implies the existence of 
a worldline of a magnetic monopole called an 't Hooft line.
By the periodicity and gauge invariance, 
we have observables such as the axion operator 
$e^{i p_\phi \phi ({\cal P})}$ and Wilson loop 
$e^{i p_a \int_{\cal C }a }$,
where the charges $p_\phi$ and $p_a$ should be
 quantized as integers.

\subsection{Invertible $\bb{Z}_N$ 0- and 1-form electric symmetries}

In this system, 
there are four kinds of invertible 
higher-form symmetries~\cite{Hidaka:2020iaz,Hidaka:2020izy}.
Two of them are associated with the equations of motion
for the axion and photon, which we will call 
electric symmetries.
Both of them are parameterized by $\bb{Z}_N$ due to the 
chiral anomaly.

By the equations of motion, 
we have the following conserved charges,
\begin{equation}
Q_\phi ({\cal V})= 
\int_{\cal V}\(-v^2 *d\phi - \fr{N}{8\pi^2} a \wed da\),
\end{equation}
and 
\begin{equation}
Q_a ({\cal S})= 
\int_{\cal S}\(\fr{1}{e^2} *da - \fr{N}{4\pi^2} \phi  da\),
\end{equation}
where ${\cal V}$ and ${\cal S}$ are 
3- and 2-dimensional closed subspaces, respectively.
These objects are topological under small deformations of 
${\cal V}$ and ${\cal S}$ thanks to the Stokes theorem and 
the equations of motion.
Therefore, we would have unitary objects given by
\begin{equation}
 U_\phi (e^{i\alpha_\phi}, {\cal V}) = \exp \(i\alpha_\phi Q_\phi ({\cal V})\), 
\quad
 U_a (e^{i\alpha_a}, {\cal S}) = \exp \(i\alpha_a Q_a({\cal S})\),
\label{221204.1335}
\end{equation}
which would be symmetry generators.
Here, the parameters would be
$\alpha_{\phi} , \alpha_a \in \bb{R}$.

However, these topological objects are not gauge invariant 
under the large gauge transformations.
The problem originates from the chiral anomaly.
The presence of the anomaly gives rise to 
$e^{\fr{ i N \alpha_\phi} {8\pi^2} \int_{\cal V} a \wed da}$
and 
$ e^{\fr{ i N \alpha_a}{4\pi^2} \int_{\cal S}\phi  da} $
in \er{221204.1335},
which depends on the gauge field $a$ and the axion $\phi$
without derivatives.
One can see this problem as follows.
To make the topological terms gauge invariant, 
we define the terms by using 
4- and 3-dimensional 
auxiliary spaces $\Omega_{\cal V}$ and ${\cal V_S}$ 
whose boundaries are ${\cal V}$ and ${\cal S}$,
i.e.,
 $ \der \Omega_{\cal V} ={\cal V}$ 
and  $ \der {\cal V_S} ={\cal S}$
as
$e^{\fr{ i N \alpha_\phi} {8\pi^2} \int_{{\Omega}_{\cal V}} d a \wed da}$
and 
$ e^{\fr{N \alpha_a}{4\pi^2} \int_{\cal V_S}d\phi \wed  da} $
respectively.
However, we have chosen the auxiliary spaces by hand.
We therefore require the absence of the ambiguity in the choice of 
the auxiliary spaces.
For other choices of the auxiliary spaces
$\Omega_{\cal V}'$ and ${\cal V_S'}$ satisfying 
$\der \Omega_{\cal V}' = {\cal V}$ and 
$\der {\cal V_S'} = {\cal S}$,
the ambiguity is absent if the following conditions are satisfied,
\begin{equation}
 e^{\fr{ i N \alpha_\phi} {8\pi^2} \int_{{\Omega}_{\cal V}} d a \wed da}
e^{- \fr{ i N \alpha_\phi} {8\pi^2} \int_{{\Omega}'_{\cal V}} d a \wed da}
=  e^{\fr{ i N \alpha_\phi} {8\pi^2} \int_{{\Omega}} d a \wed da} 
=1,
\end{equation}
and 
\begin{equation}
  e^{\fr{i N \alpha_a}{4\pi^2} \int_{\cal V_S}d\phi \wed  da}
 e^{- \fr{i N \alpha_a}{4\pi^2} \int_{\cal V_S'}d\phi \wed  da}
= 
  e^{\fr{i N \alpha_a}{4\pi^2} \int_{\cal V}d\phi \wed  da}
= 
1.
\end{equation}
Here, $\Omega = \Omega_{\cal V} \cup \b{\Omega}'_{\cal V}$ 
and ${\cal V} = {\cal V_S} \cup \b{\cal V}_{\cal S}'$
are closed 4- and 3-dimensional spaces,
where
$\b{\Omega}'_{\cal V}$ and $\b{\cal V}_{\cal S}'$
are ${\Omega}'_{\cal V}$ and ${\cal V}_{\cal S}'$ 
with opposite orientations, respectively.
By the requirements, the parameters $e^{i\alpha_\phi }$ and 
$ e^{i\alpha_a}$ are constrained as
$e^{i\alpha_\phi }, e^{i\alpha_a} \in \bb{Z}_N$.
In this restriction, the topological objects
in \er{221204.1335} give us symmetry generators of 
$\bb{Z}_N$ 0- and 1-form symmetries.
These symmetry generators are invertible because they are unitary.

Now, we can discuss symmetry transformation laws.
Charged objects of $\bb{Z}_N$ 
0- and 1-form symmetries are 
an axion two-point operator 
$e^{i\phi ({\cal P})  - i\phi ({\cal P'}) } $ 
and a Wilson loop $W({\cal C}) = e^{i\int_{\cal C} a}$.
Here, we consider $U_\phi (e^{2 \pi i /N}, {\cal V})$ 
for simplicity.
The transformation law can be expressed in terms of 
the correlation functions,
\begin{equation}
 \vevs{
U_\phi (e^{2 \pi i /N}, {\cal V})
e^{i\phi ({\cal P})  - i\phi ({\cal P'}) }
}
 = {\cal N}\int {\cal D}[\phi, a]
U_\phi (e^{2 \pi i /N}, {\cal V})
e^{i\phi ({\cal P}) - i\phi ({\cal P'}) + i S[\phi ,a]}.
\end{equation}
Here, the symbol $\vevs{...}$ denotes the vacuum expectation value
(VEV), 
${\cal D}[\phi,a]$ is an abbreviation of 
the integral measure ${\cal D} \phi {\cal D} a$,
and 
${\cal N}$ is a normalization factor such that $\vevs{1} = 1$.
We redefine the path integral variable as 
$\phi (x)\to \phi (x)+ \fr{2\pi }{N } \delta_0 (x; \Omega_{\cal V})$,
where $\delta_0 (x; \Omega_{\cal V})$ is a delta function 0-form.
For a positive integer $ 0 \leq p\leq 4$, we define 
a delta function $(4-p)$-form on a $p$-dimensional subspace $\Sigma_p$ 
as follows,
\begin{equation}
 \int_{\Sigma_p} \omega_p (x)
= \int \omega_p (x) \wed \delta_{4-p} (x; \Sigma_p) ,
\end{equation}
where $\omega_p$ is a $p$-form field.
Note that the explicit form of the delta function is 
\begin{equation}
\begin{split}
 \delta_{4-p}(x; \Sigma_{p})
&=
 \delta_{\mu_1 ... \mu_{4-p} } (x; \Sigma_{p})
dx^{\mu_1} \weds dx^{\mu_{4-p}}
\\
&
= \fr{\epsilon_{\mu_1 ... \mu_{4-p} \nu_1 ...\nu_{p}} }{p! (4-p)!}
dx^{\mu_1} \weds dx^{\mu_{4-p}}
\int_{\Sigma_p} d y^{\nu_1} \weds dy^{\nu_p}\delta^4 (x-y), 
\end{split}
\label{221207.0615}
\end{equation}
where $(y^\mu) $ represents the coordinate of $\Sigma_p$ in the spacetime.
Hereafter, we often abbreviate $\delta_{4-p} (x; \Sigma_p) $
to $\delta_{4-p} (\Sigma_p) $.
Under the redefinition, we have
the $\bb{Z}_N$ transformation of the axion operator,
\begin{equation}
\begin{split}
&
 \vevs{
U_\phi (e^{2 \pi i / N }, {\cal V})
e^{i\phi ({\cal P}) - i\phi ({\cal P'})}
}
 = 
e^{\fr{2\pi i }{N }\link ({\cal V, (P,P')}) }
 \vevs{
e^{i\phi ({\cal P}) - i\phi ({\cal P'})}
},
\end{split}
\end{equation}
where $\link ({\cal V, (P,P')}$ is 
 linking number between two points $({\cal P,P'})$
and ${\cal V}$,
\begin{equation}
 \link ({\cal V}, ({\cal P,P'}))
:= \int_{\Omega_{\cal V}} (\delta_4 ({\cal P}) - \delta_4 ({\cal P}') ).
\end{equation}

Similarly, we derive the symmetry transformation of the Wilson loop 
as follows.
In terms of the correlation function, the action of 
$U_a (e^{2 \pi i / N }, {\cal S})$ can be written as
\begin{equation}
 \vevs{
U_a (e^{2 \pi i /N}, {\cal S})
e^{i\int_{\cal C}a}
}
 = {\cal N}\int {\cal D}[\phi, a]
U_a (e^{2 \pi i /N}, {\cal S})
e^{i\int_{\cal C}a  + iS[\phi ,a]}.
\end{equation}
We redefine the path integral variable as 
$a \to a -  \fr{2\pi }{N } \delta_1 ({\cal V_S})$,
and we have
the $\bb{Z}_N$ transformation of the Wilson loop,
\begin{equation}
\begin{split}
&
 \vevs{
U_a (e^{2 \pi i / N }, {\cal S})
e^{i\int_{\cal C} a }
}
 = 
e^{\fr{2\pi i }{N }\link ({\cal S,C}) }
 \vevs{e^{i\int_{\cal C}a }},
\end{split}
\label{221204.1555}
\end{equation}
where $\link ({\cal S,C})$ is 
the linking number between ${\cal S}$ and ${\cal C}$,
\begin{equation}
 \link ({\cal S, C}) 
:= 
\int_{\cal V_S} \delta_3 ({\cal C})
=
\int \delta_3 ({\cal C}) \wed \delta_1 ({\cal V_S})
=
-\int_{\cal C} \delta_1 ({\cal V_S}).
\end{equation}
Here and hereafter, 
we assume that ${\cal S}$ does not have self intersections,
$\delta_2 ({\cal S}) \wed \delta_2 ({\cal S}) =0 $
for simplicity.

\subsection{Invertible $U(1)$ 1- and 2-form magnetic symmetries}

The rest of invertible symmetries are $U(1)$ 1- and 2-form symmetries
 associated with the Bianchi identities for the axion and photon,
$dd\phi =0$ and $dd a  =0$, respectively.
Since $dda =0$ includes the magnetic Gauss law, 
we call these symmetries magnetic symmetries. 

The Bianchi identities give us the conserved currents 
$d\phi$ and $da$, and lead to 
the following 
unitary, gauge invariant, and topological objects,
\begin{equation}
 U_{\rm 1M} (e^{i\beta_a},{\cal S})= e^{i \beta_a \int_{\cal S}\fr{da}{2\pi}  },
\qtq{and}
 U_{\rm 2} (e^{i\beta_\phi},{\cal C})= e^{i \beta_\phi 
\int_{\cal C}\fr{ d\phi}{2\pi} }.
\end{equation}
Since the integrands are gauge invariant, 
the parameters satisfy $e^{i\beta_a} ,e^{i\beta_\phi} \in U(1)$.
Note that the normalization of the integrands is determined 
by the quantization conditions of the axion and photon.

The charged objects for the magnetic 1- and 2-form symmetries 
are an 't Hooft loop and a worldsheet
of an axionic string, respectively.
For an 't Hooft loop denoted by $T(m_a, {\cal C})$
on a closed loop 
${\cal C}$ with a charge $m_a\in \bb{Z}$,  
the symmetry generator $ U_{\rm 1M} (e^{i\beta_a},{\cal S})$ 
acts on $T(m_a, {\cal C})$ as a $U(1)$ transformation,
\begin{equation}
 \vevs{ U_{\rm 1M} (e^{i\beta_a},{\cal S}) T(m_a, {\cal C})}
 = e^{i m_a \beta_a \link ({\cal S,C})} \vevs{T(m_a, {\cal C})},
\end{equation}
where we have used $\int_{\cal S} da  = 2\pi m_a \link ({\cal S,C})$ 
in the presence of $T(m_a , {\cal C})$.
Similarly, for a worldsheet of an axionic string 
denoted by $V(m_\phi, {\cal S})$
a closed 2-dimensional subspace ${\cal S}$
with a charge $m_\phi \in \bb{Z}$,
the symmetry generator 
$ U_{\rm 2} (e^{i \beta_\phi},{\cal C})$ acts on 
$V(m_\phi , {\cal S})$ as a $U(1)$ transformation,
\begin{equation}
  \vevs{ U_2 (e^{i\beta_\phi },{\cal C}) V(m_\phi, {\cal S})}
 = e^{i m_\phi \beta_\phi \link ({\cal C,S})} \vevs{V(m_\phi ,{\cal S})}.
\end{equation}

\section{Non-invertible 0-form symmetry \label{ni0}}

In this section, we review a non-invertible 0-form 
symmetry~\cite{Choi:2022jqy,Cordova:2022ieu},
which is another gauge invariant 
modification of the topological object 
$U_\phi (e^{i\alpha_ \phi }, {\cal V})$.
This modification is possible if the parameter 
$\alpha_ \phi$ is a $2\pi$ multiple of a rational number,
$\alpha_\phi \in 2\pi \bb{Q}/2\pi \bb{Z}$.
For simplicity, we consider $\alpha_\phi = 2\pi /(N q_\phi)$ with 
$q_{\phi} \in \bb{Z}$.

In this solution, we regard 
$e^{\fr{ i } { 4\pi q_\phi} \int_{\cal V} a \wed da}$
as a naive effective action of a fractional quantum Hall state,
and replace it with a path integral as
\begin{equation}
e^{\fr{i }{4\pi q_\phi} \int_{\cal V} a \wed da} 
 \to 
\int {\cal D}c \, 
e^{\fr{i }{2\pi}
\int_{\cal V} c \wed da - 
\fr{i q_\phi}{4\pi} \int_{\cal V} c \wed dc}.
\label{220514.1536}
\end{equation}
Here, $c$ is a $U(1)$ 1-form gauge field, which is defined on 
${\cal V}$.
The right-hand side represents an action of a fractional 
quantum Hall state.
We might locally integrate out $c$ as $  dc = q_\phi d a$,
and obtain the term $e^{\fr{i }{4\pi q_\phi} \int_{\cal V} a \wed da} $.
Furthermore, one can 
show that the right-hand side of \er{220514.1536}
is invariant under the choice of the auxiliary space 
$\Omega_{\cal V}$ with $\der \Omega_{\cal V} = {\cal V}$,
where $c$ is extended to $\Omega_{\cal V}$.

By applying the modification to $ U(e^{i\alpha_\phi}, {\cal V})$,
we obtain a gauge invariant and topological object,
\begin{equation}
  D_\phi (e^{2 \pi i /(N q_\phi)}, {\cal V}) 
=
 e^{i \int_{\cal V} -\fr{2\pi v^2 }{N q_\phi}*d\phi}
\int {\cal D}c \, 
e^{
\fr{i q_\phi}{4\pi} \int_{\cal V} c \wed dc
- \fr{i }{2\pi}
\int_{\cal V} c \wed da 
}.
\label{220520.2139}
\end{equation}
This object is topological because 
we can freely move the
object $D_\phi (e^{2 \pi i /(N q_\phi)}, {\cal V}) $
if the topology of ${\cal V}$ is not changed. 
We summarize the derivation in appendix~\ref{top0}.
Therefore, the defect can be thought of as a symmetry defect.
Since the defect is codimension-1, it is a 0-form symmetry defect.
Furthermore, it is a non-invertible defect
because 
the expectation value of the symmetry defect is not unity 
in general but 
depends on the topology of ${\cal V}$.

Now, we can discuss the transformation law 
of the axion under the action of non-invertible defects.
The action of the non-invertible defect on the axion two-point operator 
$e^{i\phi ({\cal P})  - i\phi ({\cal P'}) } $ is derived as follows.
We consider the following correlation function,
\begin{equation}
 \vevs{
D_\phi (e^{2 \pi i /(N q_\phi)}, {\cal V})
e^{i\phi ({\cal P})  - i\phi ({\cal P'}) }
}
 = {\cal N}\int {\cal D}[\phi, a]
D_\phi (e^{2 \pi i /(Nq_\phi)}, {\cal V})
e^{i\phi ({\cal P}) - i\phi ({\cal P'}) + iS[\phi ,a]}.
\end{equation}
and move ${\cal V}$ to another 3-dimensional subspace 
${\cal V}'$ without changing the topology of ${\cal V}$.
We will call such deformations of defects continuous deformations.
We redefine the path integral variable as 
$\phi (x)\to \phi (x)+ \fr{2\pi}{Nq_\phi} \delta_0 
(x; \Omega_{\cal V,V'})$,
and the correlation function becomes
\begin{equation}
\begin{split}
&
 \vevs{
D_\phi (e^{2 \pi i /(Nq_\phi)}, {\cal V})
e^{i\phi ({\cal P}) - i\phi ({\cal P'})}
}
 = 
e^{\fr{2\pi i }{Nq_\phi}
\link ({{\cal V}\sqcup \b{\cal V}', ({\cal P,P' })}) }
 \vevs{
D_\phi (e^{2 \pi i /(Nq_\phi)}, {\cal V}')
e^{i\phi ({\cal P}) - i\phi ({\cal P'})}
}.
\end{split}
\end{equation}
Here, 
we have introduced 
a 4-dimensional subspace $\Omega_{\cal V,V'}$ 
satisfying $\der \Omega_{\cal V,V'} = {\cal V} \sqcup \b{\cal V}'$,
where $\b{\cal V}'$ is ${\cal V}'$ with an opposite orientation. 
One of the differences in the transformation laws between the invertible and non-invertible topological defects is the possible transformation parameters.
The possible transformation parameters is constrained by $\bb{Z}_N$
for the invertible symmetry in \er{221204.1555},
but
we can take an arbitrary small transformation parameter 
$\fr{2\pi }{Nq_\phi}$ for the non-invertible symmetry.

\section{Non-invertible 1-form symmetry \label{ni1}}

In this section, we argue that there is a non-invertible 1-form symmetry 
parameterized by a rational number.
First, we modify the problematic term 
$e^{\fr{i N \alpha_a }{4\pi^2}\int_{\cal S} \phi da }$ 
of $U_a (e^{i\alpha_a}, {\cal S})$ in \er{221204.1335}
by introducing additional dynamical fields defined on ${\cal S}$
such that it becomes gauge invariant.
We then consider the symmetry transformation law of the 
Wilson loop, which gives us a phase rotation by a rational number.

\subsection{Non-invertible symmetry defect}

Here, we consider the modification of 
$e^{\fr{i N \alpha_a }{4\pi^2}\int_{\cal S} \phi da }$ 
in $U_a(e^{i\alpha_a}, {\cal S})$.
For simplicity, 
we consider the case of 
$\alpha_a = 2\pi /(N q_a)$ with $q_a \in \bb{Z}$.
We regard $e^{\fr{i N \alpha_a }{4\pi^2}\int_{\cal S} \phi da }$ 
as a naive effective action of a TQFT,
and find a gauge invariant expression of it 
in terms of additional degrees of freedom 
defined on ${\cal S}$.

We argue that we can make the topological object 
$U_a(e^{2\pi i/ (N q_a)}, {\cal S})$ gauge invariant 
by the following modification,
\begin{equation}
 e^{\fr{i }{2\pi q_a} \int_{\cal S} \phi  da}
\to 
\int {\cal D}[\chi, u]
e^{ \fr{i}{2\pi } \int_{\cal S} ( \chi da
+u \wed d\phi 
- q_a \chi du)}
.
\label{221204.1628}
\end{equation}
Here, 
$\chi$ is a $2\pi$ periodic scalar, and $u$ is a $U(1)$ 
1-form gauge field quantized as $\int_{\cal S} du \in 2\pi \bb{Z}$.
Both are defined on the 2-dimensional space ${\cal S}$.
We find that the TQFT in the 
right-hand side represents a low-energy effective action 
of charge $q_a$ Schwinger model~\cite{Coleman:1974bu,Armoni:2018bga,Misumi:2019dwq}.
The right-hand side does not depend on the choice of a
3-dimensional auxiliary space
with an extension of $\chi$ and $u$ to the auxiliary space
because of the quantization conditions 
$e^{ \fr{i}{2\pi } \int_{\cal V} d \chi \wed da} =
e^{\fr{i}{2\pi} \int_{\cal V}  d\phi \wed du} 
= e^{\fr{i q_a}{2\pi} \int_{\cal S} d\chi \wed du} = 1 $
for a closed 3-dimensional space.
One can check that we may locally 
have $ e^{\fr{i }{2\pi q_a} \int_{\cal S} \phi  da}$
by eliminating $u$ and $\chi$ using their equations of motion,
 $d\phi = q_a d\chi$ and $ da = q_a du$. 

By this modification, we have a gauge invariant and topological object as  
\begin{equation}
D_a (e^{2 \pi i /(N q_a)}, {\cal S}) 
= 
e^{\fr{2 \pi i }{ N q_a} \int_{\cal S} \fr{1}{e^2}*da}
\int {\cal D} [\chi,u] 
e^{ \fr{i}{2\pi } \int_{\cal S} 
(q_a \chi du - \chi da - u \wed d\phi   ) 
}.
\label{220721.2206}
\end{equation}
The property of $D_a (e^{2 \pi i /(N q_a)}, {\cal S}) $ 
under a continuous deformation of ${\cal S}$
 is shown in appendix~\ref{top1}.
Since it is a codimension-2 defect, 
it is a 1-form symmetry defect.
The symmetry described by $D_a (e^{2 \pi i /(N q_a)}, {\cal S}) $ is non-invertible because the partition function of the TQFT is
not unity but depends on 
the topology of ${\cal S}$ in general.

\subsection{Transformation law}
 
We discuss the transformation laws for a Wilson loop.
We consider the correlation function,
\begin{equation}
 \vevs{D_a (e^{2\pi i /(Nq_a)},{\cal S}) e^{i \int_{\cal C} a}}
= {\cal N}
\int {\cal D}[\phi ,a ] 
D_a (e^{2\pi i /(Nq_a)},{\cal S})
e^{iS [\phi,a] + i \int_{\cal C} a }. 
\end{equation} 
As in the case of the 0-form symmetry acting on the axion, 
we continuously deform ${\cal S}$ to 
another 2-dimensional subspace ${\cal S}'$ 
by the redefinition of the dynamical field 
$a \to a - \fr{2\pi }{Nq_a} \delta_1 ({\cal V}_{\cal S,S'})$,
and we have
\begin{equation}
   \vevs{
D_a (e^{2\pi i p_a/(Nq_a)},{\cal S}) e^{iq \int_{\cal C} a}
}
= 
e^{\fr{2\pi i }{N q_a} \link ({\cal S}\sqcup \b{\cal S}', {\cal C})}
\vevs{D_a (e^{2\pi i p_a/(Nq_a)},{\cal S}')
e^{ i \int_{\cal C} a }}.
\end{equation}
Here, we have introduced a 3-dimensional subspace 
${\cal V}_{\cal S,S'}$
satisfying $\der {\cal V}_{\cal S,S'} = {\cal S}\sqcup \b{\cal S}'$.
Therefore, 
we obtain a symmetry transformation of the Wilson loop 
parameterized by a rational number $1/(N q_a)$, which 
can have many parameter than $\bb{Z}_N $. 

Before going to the next section, we comment on some differences 
between invertible and non-invertible 1-form symmetries.
In the case of the 1-form invertible symmetry, 
it is parameterized by the finite number $\bb{Z}_N$,
and it becomes trivial for the simplest case $N =1$.
Meanwhile, in the case of the non-invertible 1-form symmetry,
it is parameterized by a rational number $1/(N q_a)$.
We can take infinitely many parameters, 
and the parameter can be arbitrarily small.
Furthermore, the non-invertible 1-form symmetry is meaningful 
even for the simplest case $N = 1$.

\section{Correlation between 0- and 1-form non-invertible defects\label{corr}}

We now discuss the correlation functions 
between non-invertible defects.
We show that 
an  intersection of non-invertible 0- and 1-form symmetry defects leads to a magnetic 1-form 
symmetry defect.
Further, an intersection of two non-invertible 1-form symmetry defects induces a magnetic 2-form symmetry defect.
Both magnetic symmetry defects have boundaries on 
the intersections.
We then show that these intersections are charged under 
the non-invertible symmetries.

\subsection{Intersection of non-invertible 0- and 1-form symmetry defects
\label{int01}}
\begin{figure}[t]
\begin{center}
  \ig[height=10em]{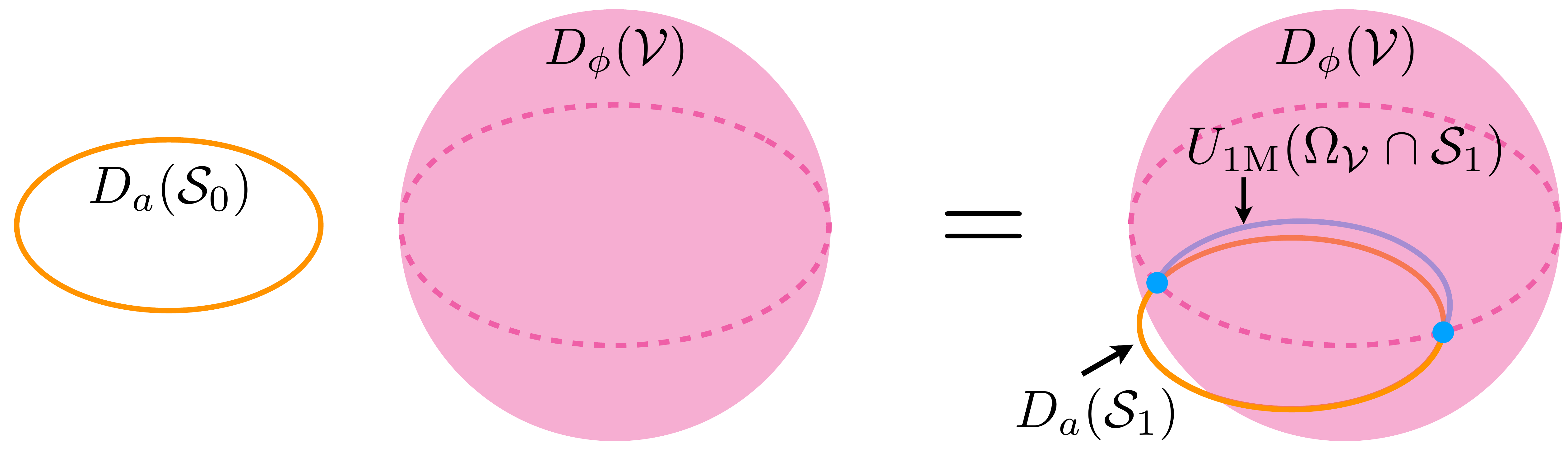}
\end{center}
\caption{\label{D01}
Configuration of symmetry defects in \er{221207.1422}.
This figure is a time slice of the configuration.
Non-invertible 
0- and 1-form symmetry defects are expressed by pink surfaces  
and orange lines.
The invertible magnetic 1-form symmetry generator is displayed as 
a blue line.
We have omitted the parameters $e^{2\pi i /(N q_\phi)}$ 
and $e^{2\pi i /(N q_a)}$ of symmetry defects.}
\end{figure}

We first consider the following correlation function,
\begin{equation}
\vevs{D_\phi  (e^{2 \pi i /(N q_\phi)}, {\cal V}) 
D_a (e^{2 \pi i /(N q_a)}, {\cal S}_0)}.
\end{equation}
At this stage, we assume that ${\cal V}$ and
${\cal S}_0 $ do not have intersections.
Therefore, the correlation function is reduced to 
a finite value that depends on the topology of ${\cal S}_0 $ 
and ${\cal V}$ in general. 
Now, we continuously deform ${\cal S}_0$ to 
${\cal S}_1 $, 
where ${\cal S}_1 $ intersects with ${\cal V}$
as shown in Fig.~\ref{D01}.
This can be done by taking
a 3-dimensional subspace ${\cal V}_{01}$
such that 
$\der {\cal V}_{01} = {\cal S}_0 \sqcup \b{\cal S}_1$,
where $\b{\cal S}_1$ is ${\cal S}_1$ with an opposite orientation.
By the deformation, the photon receives the 
redefinition 
$a \to 
a
- \fr{2\pi }{N q_a } \delta_1 ({\cal V}_{01})$
as shown in Appendix \ref{top1},
\begin{equation}
\begin{split}
&
\vevs{
D_\phi  (e^{2 \pi i /(N q_\phi)}, {\cal V}) 
D_a (e^{2 \pi i /(N q_a)}, {\cal S}_0)
}
\\
&=
{\cal N} 
\int {\cal D}[\phi, a] 
D_a (e^{2 \pi i /(N q_a)}, {\cal S}_1)
e^{iS[\phi, a]} e^{i \int_{\cal V} -\fr{2\pi v^2 }{N q_\phi}*d\phi}
\\
&
\quad
\times
\int {\cal D}c \, 
e^{ \fr{i }{ N q_a}
\int_{\Omega_{\cal V}} d c \wed d \delta_1 ({\cal V}_{01})}
e^{
\fr{i q_\phi}{4\pi} \int_{\cal V} c \wed dc
- \fr{i }{2\pi}
\int_{\cal V} c \wed da 
}.
 \end{split}
\label{221207.1422}
\end{equation}
Here, we have taken an extension of $c$ on ${\cal V}$ to $\Omega_{\cal V}$.
We thus have an object
$e^{ \fr{i }{ N q_a}
\int_{\Omega_{\cal V}} d c \wed d \delta_1 ({\cal V}_{01})}$
from 
$D_\phi  (e^{2 \pi i /(N q_\phi)}, {\cal V}) $.
We can further rewrite the correlation function 
by the field redefinition 
$ c + \fr{2\pi }{N q_a q_\phi}  \delta_1 ({\cal V}_{01}) \to c $ 
on $\Omega_{\cal V}$
as
\begin{equation}
\begin{split}
&
\vevs{
D_\phi  (e^{2 \pi i /(N q_\phi )}, {\cal V}_0) 
D_a (e^{2 \pi i/(N q_a)}, {\cal S})}
\\
&
=
\vevs{
U_{\rm 1M}
(e^{\fr{2\pi i }{N q_\phi q_a}},\Omega_{\cal V} \cap {\cal S}_1)
D_\phi  (e^{2 \pi i /(N q_\phi )}, {\cal V})  
D_a (e^{2 \pi i/(N q_a)}, {\cal S}_1)}.
 \end{split}
\end{equation}
We find that the correlation function leads to the magnetic 1-form symmetry 
generator 
$U_{\rm 1M}
(e^{\fr{2\pi i }{N q_\phi q_a}},\Omega_{\cal V} \cap {\cal S}_1)
 = e^{\fr{i }{N q_\phi q_a} \int_{\Omega_{\cal V} \cap {\cal S}_1} da }$ 
that has a 1-dimensional boundary, 
$\der( \Omega_{\cal V} \cap {\cal S}_1) = {\cal V} \cap {\cal S}_1$.
The magnetic symmetry generator would be thought as 
a worldline of a fractionally charged electric particle
$e^{\fr{i }{N q_\phi q_a} \int_{{\cal V}_1  \cap {\cal S}} a }$,
but it violates the quantization condition.
Therefore, this fractionally charged object 
should be treated as an object on the boundary of the surface 
$\Omega_{\cal V} \cap {\cal S}_1 $.

\subsection{Intersection of two non-invertible 1-form symmetry defects}

\begin{figure}[t]
\begin{center}
 \ig[height=10em]{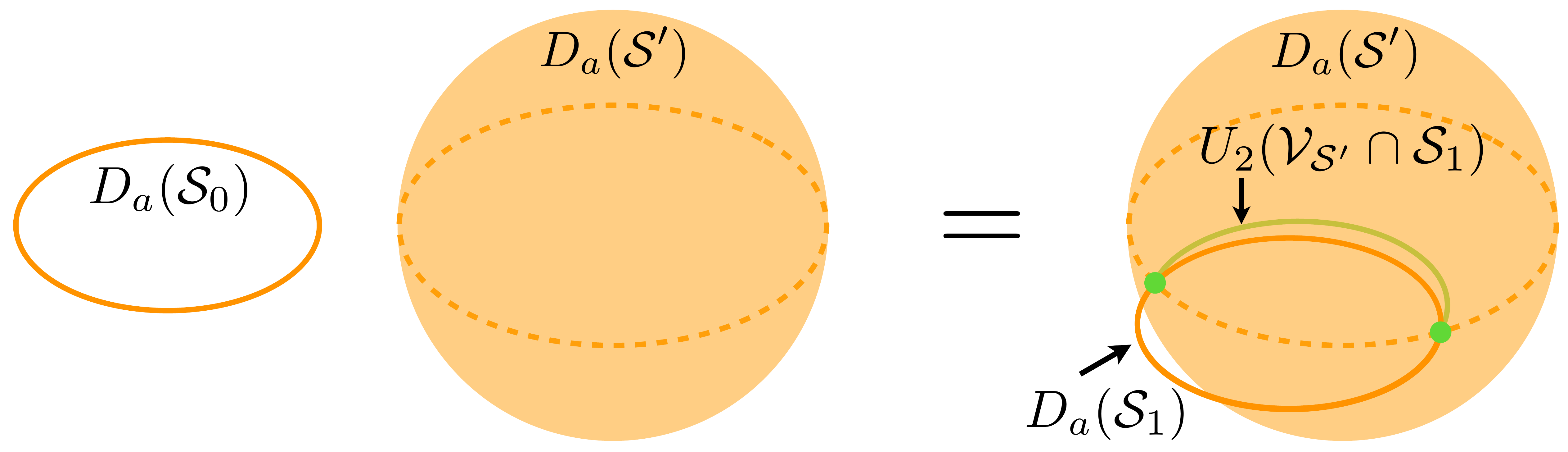} 
\end{center}
\caption{\label{D11} 
Configuration of symmetry defects in \er{221207.1422}.
This figure is a time slice of the defects.
Non-invertible 
1-form symmetry defects are expressed by 
and orange lines or surfaces.
The invertible magnetic 2-form symmetry generator 
is displayed as a green line.
We have omitted the parameters of symmetry defects.}
\end{figure}

Next, we discuss intersections of two non-invertible 1-form 
symmetry defects.
We begin with a correlation function 
of two 1-form symmetry defects,
\begin{equation}
\vevs{D_a (e^{2 \pi i /(N q_a)}, {\cal S}_0)
D_a (e^{2 \pi i /(N q'_a)}, {\cal S})},
\end{equation}
where ${\cal S}_0 $ and $ {\cal S}$ do not have intersections,
${\cal S}_0 \cap {\cal S} = \emptyset$.
This correlation function again reduces to a finite value.
Now, we continuously deform $ {\cal S}_0$ to ${\cal S}_1$, and 
has intersections with ${\cal S}$ satisfying 
$\delta_2 ({\cal S}_1) \wed \delta_2 ({\cal S}) \neq 0$
(see Fig.~\ref{D11}).
The deformation can be done by using a 3-dimensional subspace 
${\cal V}_{10}$ with the boundary 
$\der {\cal V}_{01} = {\cal S}_0 \sqcup \b{\cal S}_1$.
By using a similar procedure in section \ref{int01},
we have the following correlation function, 
\begin{equation}
\begin{split}
&
\vevs{
D_a (e^{2 \pi i /(N q_a)}, {\cal S}_0)
D_a (e^{2 \pi i /(N q'_a)}, {\cal S})
}
\\
&=
{\cal N} 
\int {\cal D}[\phi, a] 
D_a (e^{2 \pi i /(N q_a)}, {\cal S}_1)
e^{iS[\phi, a] + 
\fr{2 \pi i }{ N q_a} \int_{\cal S} \fr{1}{e^2}*da}
\\
&
\quad
\times
\int {\cal D} [\chi,u] 
e^{ 
\fr{i}{ Nq_a} \int_{\cal V_S} 
 d\chi \wed  
 d \delta_1 ({\cal V}_{01}) 
}
e^{ 
\fr{i}{2\pi } \int_{\cal S} 
(q'_a \chi du - \chi da -  u \wed d\phi   ) 
}
\\
&=
\vevs{U_2 (e^{\fr{2\pi i }{Nq_a q_a'}}, {\cal V}_{\cal S} \cap {\cal S}_1)
D_a (e^{2 \pi i /(N q_a)}, {\cal S}_1)
D_a (e^{2 \pi i /(N q_a')}, {\cal S})}
 \end{split}
\label{221207.1423}
\end{equation}
where we have redefined the variables 
$a \to a - \fr{2\pi }{Nq_a} \delta_1 ({\cal V}_{01})$
and then $u \to u - \fr{2\pi }{Nq_a q_a'} \delta_1 ({\cal V}_{01}) $
on ${\cal V_S}$.
By the correlation function, we have
the 2-form symmetry generator 
$U_2 (e^{\fr{2\pi i }{Nq_a q_a'}}, {\cal V}_{\cal S}  \cap {\cal S}_1)
 = 
e^{ 
\fr{i}{Nq_a q_a'} 
\int_{ {\cal V}_{\cal S} \cap {\cal S}_1} 
 d\phi 
}
$
with 0-dimensional boundaries,
$\der  ({\cal V}_{\cal S} \cap {\cal S}_1 ) 
=  {\cal S} \cap  {\cal S}_1$.
This 2-form symmetry generator could be thought of as an 
axion operator 
with a fractional charge
$e^{ \fr{i}{Nq_a q_a'} \phi ({\cal S} \cap {\cal S}_1)}$, 
but it violates the 
$2\pi$ periodicity of the axion.
To preserve the periodicity, this axion operator should 
be defined as a boundary of the 1-dimensional subspace
$ {\cal V}_{\cal S} \cap {\cal S}_1$. 

\subsection{Detecting fractional charges by non-invertible defects\label{corr3}}

In the above discussion, we have considered the correlation of 
two non-invertible defects.
We have found that there are induced invertible 1- and 2-form 
magnetic symmetries with boundaries by intersecting
non-invertible symmetry defects, which could be thought as 
fractionally charged Wilson loops and axion operators, 
respectively.
Here, we show that 
the intersecting non-invertible defects 
have charged objects of the non-invertible 
1- and 0-form symmetries,
discuss their physical meaning of them.

\begin{figure}[t]
\begin{center}
 \ig[height=10em]{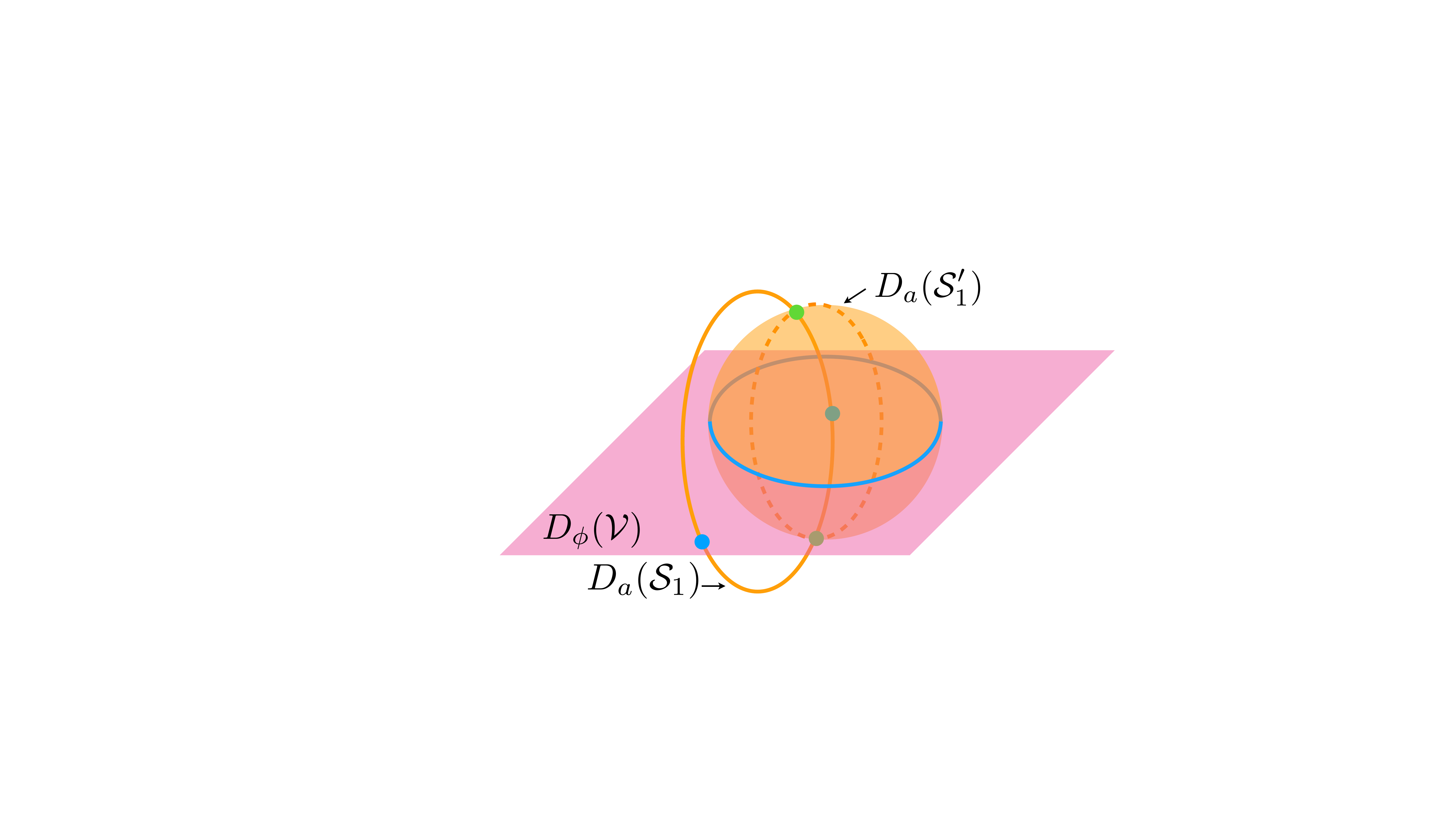} 
\end{center}
\caption{\label{D011} Configuration of symmetry defects in 
the right-hand side of \er{221207.0546}.
A time slice of the configuration is shown.
Non-invertible 
0- and 1-form symmetry defects are colored
pink and orange, respectively.
We only display the boundaries of the magnetic symmetry defects 
for simplicity.
The blue lines and dots represent the boundaries of magnetic 1-form 
symmetry defects.
The green dots correspond to the boundaries of magnetic 2-form 
symmetry defect.
The parameters of symmetry defects have been omitted.
}
\end{figure}

First, we show that the intersection of non-invertible 0- and 1-form symmetry generators is charged under 
the non-invertible 1-form symmetry.
To see this, we consider the following 
correlation function, 
\begin{equation}
 \vevs{
D_a (e^{2 \pi i /(N q_a)}, {\cal S}_0)
D_a (e^{2 \pi i /(N q'_a)}, {\cal S}'_0)
D_\phi  (e^{2 \pi i /(N q_\phi)}, {\cal V}) 
},
\end{equation}
where neither ${\cal S}_0$, ${\cal S}'_0 $, nor ${\cal V}$
have intersections.
This correlation function is again reduced to 
a finite value depending on the topology of the defects
in general.
Now, we continuously 
deform ${\cal S}_0$ to ${\cal S}_1$.
Here, we take ${\cal S}_1$
such that it intersects with ${\cal V}$.
By using the same procedure as that of section \ref{int01}
with the 3-dimensional subspace ${\cal V}_{01}$,
we have the induced magnetic 1-form symmetry defect with
the boundary ${\cal V} \cap {\cal S}_1$,
\begin{equation}
\begin{split}
&   \vevs{
D_a (e^{2 \pi i /(N q_a)}, {\cal S}_0)
D_a (e^{2 \pi i /(N q'_a)}, {\cal S}'_0)
D_\phi  (e^{2 \pi i /(N q_\phi)}, {\cal V}) 
}
\\
&
 = 
  \vevs{
U_{\rm 1M}
(e^{\fr{2\pi i }{N q_\phi q_a}},\Omega_{\cal V} \cap {\cal S}_1)
D_a (e^{2 \pi i /(N q_a)}, {\cal S}_1)
D_a (e^{2 \pi i /(N q'_a)}, {\cal S}'_0)
D_\phi  (e^{2 \pi i /(N q_\phi)}, {\cal V}) 
}.
\end{split}
\end{equation}
We then continuously deform ${\cal S}'_0 $ to ${\cal S}'_1$,
where ${\cal S}'_1$ intersects with ${\cal V}$ and ${\cal S}_1$
(see Fig.~\ref{D011}).
The deformation is given by the 3-dimensional subspace 
${\cal V}_{01}'$ with the boundary, 
$\der {\cal V}'_{01} = {\cal S}'_0 \sqcup \b{\cal S}'_1$.
The correlation function becomes
\begin{equation}
\begin{split}
&   \vevs{
D_a (e^{2 \pi i /(N q_a)}, {\cal S}_0)
D_a (e^{2 \pi i /(N q'_a)}, {\cal S}'_0)
D_\phi  (e^{2 \pi i /(N q_\phi)}, {\cal V}) 
}
\\
&
 = 
e^{- \fr{2\pi i }{N^2 q_\phi q_a' q_a} 
\link ({\cal S}'_1, {\cal V} \cap {\cal S}_1)}
\\
&
\quad
\times
  \vevs{
U_{\rm 1M}
(e^{\fr{2\pi i }{N q_\phi q_a}},\Omega_{\cal V} \cap {\cal S}_1)
U_{\rm 1M}
(e^{\fr{2\pi i }{N q_\phi q_a'}},\Omega_{\cal V} \cap {\cal S}'_1)
U_{2}
(e^{\fr{2\pi i }{N q_a q_a'}},{\cal V}_{{\cal S}_1 } \cap {\cal S}_1')
\\
&
\qquad\quad
\times
D_a (e^{2 \pi i /(N q_a)}, {\cal S}_1)
D_a (e^{2 \pi i /(N q'_a)}, {\cal S}'_1)
D_\phi  (e^{2 \pi i /(N q_\phi)}, {\cal V}) 
}.
\end{split}
\label{221207.0546}
\end{equation}
By the deformations, 
we have a fractional phase $e^{\fr{2\pi i }{N^2 q_\phi q_a' q_a} 
\link ({\cal S}'_1, {\cal V}\cap {\cal S}_1)}$
in addition to magnetic symmetry generators with boundaries.
Here, the linking number $\link ({\cal S}'_1, {\cal V}\cap {\cal S}_1)$ 
is 
given by
\begin{equation}
\link ({\cal S}'_1, {\cal V}\cap {\cal S}_1)
=
\int_{{\cal V}_{{\cal S}'_1}} 
\delta_1 ({\cal V}) \wed \delta_2({\cal S}_1)
= 
\int 
\delta_1 ({\cal V}) \wed \delta_2({\cal S}_1)
\wed \delta_1 ({\cal V}_{{\cal S}'_1})
\end{equation}
This fractional phase implies that 
the non-invertible 1-form symmetry generator 
$D_a (e^{2 \pi i /(N q_a')}, {\cal S}'_1)$
detects the fractional charge $\fr{1}{N q_\phi q_a}$
localized on the intersection 
${\cal V} \cap {\cal S}_1$,
because we can equivalently rewrite the relation in \er{221207.0546} 
as
\begin{equation}
\begin{split}
&
  \vevs{
U_{\rm 1M}
(e^{\fr{2\pi i }{N q_\phi q_a}},\Omega_{\cal V} \cap {\cal S}_1)
U_{\rm 1M}
(e^{\fr{2\pi i }{N q_\phi q_a'}},\Omega_{\cal V} \cap {\cal S}'_1)
U_{2}
(e^{\fr{2\pi i }{N q_a q_a'}},{\cal V}_{{\cal S}_1 } \cap {\cal S}_1')
\\
&
\qquad\quad
\times
D_a (e^{2 \pi i /(N q_a)}, {\cal S}_1)
D_a (e^{2 \pi i /(N q'_a)}, {\cal S}'_1)
D_\phi  (e^{2 \pi i /(N q_\phi)}, {\cal V}) 
}
\\
&
=
e^{\fr{2\pi i }{N^2 q_\phi q_a' q_a} 
\link ({\cal S}'_1, {\cal V} \cap {\cal S}_1)}
\\
&
\quad 
\times
  \vevs{
U_{\rm 1M}
(e^{\fr{2\pi i }{N q_\phi q_a}},\Omega_{\cal V} \cap {\cal S}_1)
D_a (e^{2 \pi i /(N q_a)}, {\cal S}_1)
D_a (e^{2 \pi i /(N q'_a)}, {\cal S}'_0)
D_\phi  (e^{2 \pi i /(N q_\phi)}, {\cal V}) 
}.
\end{split}
\label{221231.1821}
\end{equation}

Second, we intersect 1-form symmetry defects
$D_a (e^{2 \pi i /(N q_a)}, {\cal S}_0)$
with 
$D_a (e^{2 \pi i /(N q'_a)}, {\cal S}'_0)$
by a continuous deformation of ${\cal S}_0$ to ${\cal S}_1$,
and then intersect 
the 0-form symmetry defect 
$D_\phi  (e^{2 \pi i /(N q_\phi)}, {\cal V}) $
with ${\cal S}_1$ and ${\cal S}_0'$
 by a continuous deformation of ${\cal V}$ to ${\cal V}_1$.
In this case, 
we again obtain a fractional phase
\begin{equation}
\begin{split}
&   \vevs{
D_a (e^{2 \pi i /(N q_a)}, {\cal S}_0)
D_a (e^{2 \pi i /(N q'_a)}, {\cal S}'_0)
D_\phi  (e^{2 \pi i /(N q_\phi)}, {\cal V}) 
}
\\
&
=
\vevs{
U_2 (e^{\fr{2\pi i }{Nq_a q_a'}},  {\cal V}_{{\cal S}_0}'\cap {\cal S}_1)
D_a (e^{2 \pi i /(N q_a)}, {\cal S}_1)
D_a (e^{2 \pi i /(N q_a')}, {\cal S}'_0)
D_\phi  (e^{2 \pi i /(N q_\phi)}, {\cal V}) 
}
\\
&
 = 
e^{-\fr{2\pi i }{N^2 q_\phi q_a' q_a} 
\link ({\cal V}_1, {\cal S}_1 \cap {\cal S}'_0)}
\\
&
\quad
\times
  \vevs{
U_2 (e^{\fr{2\pi i }{Nq_a q_a'}},  {\cal V}_{{\cal S}_0}'\cap {\cal S}_1)
U_{\rm 1M}
(e^{\fr{2\pi i }{N q_\phi q_a}}, {\cal V}_{{\cal S}_1} \cap {\cal V}_1)
U_{\rm 1M}
(e^{\fr{2\pi i }{N q_\phi q_a'}},{\cal V}_{{\cal S}_0'} \cap {\cal V}_1)
\\
&
\qquad\quad
\times
D_a (e^{2 \pi i /(N q_a)}, {\cal S}_1)
D_a (e^{2 \pi i /(N q'_a)}, {\cal S}'_0)
D_\phi  (e^{2 \pi i /(N q_\phi)}, {\cal V}_1) 
}.
\end{split}
\label{221207.0712}
\end{equation}
Here, the linking number $\link ({\cal V}_1, {\cal S}_1 \cap {\cal S}'_0)$ 
is defined by
\begin{equation}
\link ({\cal V}_1, {\cal S}_1 \cap {\cal S}'_0)
=
\int_{\Omega_{{\cal V}_1} } 
\delta_2 ({\cal S}_1) \wed \delta_2({\cal S}'_0)
=
\int 
\delta_2 ({\cal S}_1) \wed \delta_2({\cal S}'_0)
\wed \delta_0 (\Omega_{{\cal V}_1}),
\end{equation}
and $\Omega_{{\cal V}_1} $ is a 4-dimensional subspace whose 
boundary is ${\cal V}_1$.
This correlation function means that 
the non-invertible 0-form symmetry defect captures the 
axion operator localized on ${\cal S}_1 \cap {\cal S}'_0$
with the fractional charge $\fr{1}{N q_a q_a'}$.
This property can be explicitly seen by 
rewriting \er{221207.0712} as
\begin{equation}
\begin{split}
&
  \vevs{
U_2 (e^{\fr{2\pi i }{Nq_a q_a'}},  {\cal V}_{{\cal S}_0}'\cap {\cal S}_1)
U_{\rm 1M}
(e^{\fr{2\pi i }{N q_\phi q_a}}, {\cal V}_{{\cal S}_1} \cap {\cal V}_1)
U_{\rm 1M}
(e^{\fr{2\pi i }{N q_\phi q_a'}},{\cal V}_{{\cal S}_0'} \cap {\cal V}_1)
\\
&
\qquad
\times
D_a (e^{2 \pi i /(N q_a)}, {\cal S}_1)
D_a (e^{2 \pi i /(N q'_a)}, {\cal S}'_0)
D_\phi  (e^{2 \pi i /(N q_\phi)}, {\cal V}_1) 
}
\\
&
=
e^{\fr{2\pi i }{N^2 q_\phi q_a' q_a} 
\link ({\cal V}_1, {\cal S}_1 \cap {\cal S}'_0)}
\\
&
\quad
\times 
\vevs{
U_2 (e^{\fr{2\pi i }{Nq_a q_a'}},  {\cal V}_{{\cal S}_0}'\cap {\cal S}_1)
D_a (e^{2 \pi i /(N q_a)}, {\cal S}_1)
D_a (e^{2 \pi i /(N q_a')}, {\cal S}'_0)
D_\phi  (e^{2 \pi i /(N q_\phi)}, {\cal V}) 
}
.
\end{split}
\label{221231.1822}
\end{equation}

\subsection{Physical meaning of correlation functions\label{phys}}

In the above discussions, we have obtained the correlation 
functions of symmetry generators. 
Now, we discuss the physical meaning of the correlation functions.

First, we can understand the 0-form symmetry defect 
$D_\phi  (e^{2 \pi i /(N q_\phi)}, {\cal V}) $ as 
a worldvolume of an axionic domain wall.
This can be seen as a solution to the equation of motion for the axion in the presence of 
the symmetry defect $d\phi  = \fr{2\pi}{N q_\phi } \delta_1 ({\cal V}) $ 
after turning off the photon,
and integrate out the gauge field $c$ on ${\cal V}$.
By using the explicit form of the delta function in \er{221207.0615}, 
we can see that $\der_i \phi =  \fr{2\pi}{N q_\phi } \delta_i ({\cal V}) $
for temporally extended ${\cal V}$. 
Similarly, we can regard the 1-form symmetry defect 
$D_a (e^{2 \pi i /(N q_a)}, {\cal S})$
as a worldsheet of a magnetic flux tube or an instantaneous surface of 
electric flux.
This can again be seen as a solution to 
the equation of motion for the photon in the presence 
of the 1-form symmetry defect, 
$da = \fr{2\pi }{N q_a } \delta_2 ({\cal S}) $.
The configuration gives us the localized magnetic 
flux $f_{ij} = \fr{2\pi}{Nq_a } \delta_{ij} ({\cal S})$
by taking ${\cal S}$ as a temporally extended surface,
or the instantaneous electric flux
$f_{0i} = \fr{2\pi}{Nq_a } \delta_{0i} ({\cal S})$ by taking 
${\cal S}$ as a temporally localized surface.
Here, $f_{\mu\nu} = \der_\mu a_\nu - \der_\nu a_\mu$ is the field strength of the photon.
Note that the directions of the magnetic and electric flux 
are parallel and perpendicular to the time slice of the surfaces,
respectively.

Now we discuss the physical meaning of correlation functions 
in \ers{221231.1821} and \eqref{221231.1822}.
First, the correlation function in \er{221231.1821} implies 
the Sikivie effect.
The Sikivie effect means that there is an induced electric charge at the intersection of the axionic domain wall and magnetic flux tube.
This is due to the modification of the electric Gauss law 
by the axion, 
$\fr{1}{e^2}\na \cdot \bs{E} = \fr{N}{4\pi^2 } \na \phi \cdot \bs{B}$.
In our case, we take ${\cal V}$ and ${\cal S}_1$ as a temporally 
extended subspaces.
By substituting the above 
configuration of the axion and photon
to the electric Gauss law, 
we have the induced electric charge
$\fr{N}{4\pi^2 } \cdot \fr{2\pi}{N q_\phi} \cdot \fr{2\pi}{N q_a} 
= \fr{1 }{N q_\phi q_a}$.
The charge coincides with the fractional charge detected by 
$D_a (e^{2 \pi i /(N q_a')}, {\cal S}'_1)$ of 
the correlation 
function in \er{221231.1821}
by taking ${\cal S}'_1$ as 
a temporally localized surface so that it links with the 
intersection ${\cal V} \cap {\cal S}_1$.

Second, by taking another choice of defects, 
we can further relate the correlation function in \er{221231.1821}
to the anomalous Hall effect on the domain wall:
there is an induced electric current on 
the axionic domain wall by adding electric flux,
which is due to the modification of the Maxwell-Amp\`ere law,
$\fr{1}{e^2}(\na \times \bs{B} -\pd{\bs{E}}{t} )
= \fr{N}{4\pi^2} (\na \phi \times \bs{E} + \bs{B}\pd{\phi}{t} ) $.
In our case, we take ${\cal V}$ and ${\cal S}_1$ as a temporally 
extended and localized subspaces, respectively.
In this choice, there is an induced instantaneous current
on the intersection ${\cal V} \cap {\cal S}_1$.
By substituting the above 
configuration of the axion and photon
to the Maxwell-Amp\`ere law, 
we have the induced electric current with the charge
$\fr{N}{4\pi^2 } \cdot \fr{2\pi}{N q_\phi} \cdot \fr{2\pi}{N q_a} = \fr{1 }{N q_\phi q_a}$. 
The value of the current again coincides with the value detected by 
$D_a (e^{2 \pi i /(N q'_a)}, {\cal S}'_1)$ 
of 
the correlation 
function in \er{221231.1821}
by taking ${\cal S}'_1$ as 
a temporally extended surface so that it links with the 
intersection ${\cal V} \cap {\cal S}_1$.

Third, the correlation function in \er{221231.1822} can be 
understood as an induced axion on the intersection of 
magnetic and electric flux.
This can be seen by the equation of motion for the axion
$v^2 \der^2 \phi = \fr{N}{4\pi^2} \bs{E} \cdot \bs{B}$,
which means that $\bs{E} \cdot \bs{B}$ is a source of the axion.
The intersection of the electric and magnetic flux
can be realized by taking ${\cal S}_1$ and ${\cal S}_0'$
as temporally extended and localized surfaces.
In this choice, $D_a (e^{2 \pi i /(N q_a)}, {\cal S}_1)$
and 
$D_a (e^{2 \pi i /(N q'_a)}, {\cal S}'_0)$ represent 
magnetic and electric flux 
with $f_{ij} = \fr{2\pi}{N q_a} \delta_{ij} ({\cal S}_1)$
and $f_{0i} = \fr{2\pi}{N q_a'} \delta_{0i} ({\cal S}'_0)$,
respectively.
By substituting the configuration, 
the charge of the induced axion is 
$ \fr{N}{4\pi^2} \cdot  \fr{2\pi}{N q_a} \cdot \fr{2\pi}{N q_a'}
 =   \fr{1}{N q_a q_a'}$,
which coincides with the fractional charge detected by 
$D_\phi  (e^{2 \pi i /(N q_\phi)}, {\cal V}_1) $ with a
temporally extend worldvolume 
${\cal V}_1$ linked with ${\cal S}_1\cap {\cal S}'_0$.

We discuss a difference 
of the discussions on the above effects 
between the invertible and non-invertible symmetries.
In the case of the invertible symmetries, 
we can discuss correlation functions with the same configurations 
as those of \ers{221231.1821} and \eqref{221231.1822} 
(see e.g., Refs.~\cite{Hidaka:2021mml,Hidaka:2021kkf}).
We can also relate the correlation functions to the Sikivie effects,
anomalous Hall effects and the induced axion.
However, the possible values of the axionic domain walls,
electric flux and magnetic flux are constrained by $\bb{Z}_N$
if we use invertible symmetry generators.
In particular, it is difficult to take an arbitrarily small value
of, e.g.,~the magnetic flux. 
In the case of non-invertible symmetries, 
this constraint is now resolved
because we can take arbitrarily small parameters
$1/(N q_\phi)$, $1/(N q_a)$, and $1/(N q_a')$.
Therefore, the non-invertible symmetries give us a more refined understanding of the 
above effects in terms of global symmetries than 
the previous understandings based on the invertible symmetries.

Before closing this section, we comment on a relationship between
the correlation functions and an 't~Hooft anomaly.
The correlation functions obtained in \ers{221231.1821} 
and \eqref{221231.1822} means that the intersections of symmetry defects become charged objects with fractional charges.
These correlation functions 
may mean a mixed 't Hooft anomaly between non-invertible 
0- and 1-form symmetries.
In the case of the invertible symmetries, it can be understood 
as an 't Hooft anomaly by using the method of the 
background gauging of higher-form symmetries~\cite{Hidaka:2020izy}
and it is called a 2-group anomaly~\cite{Benini:2018reh}.
Our correlation functions may imply a generalization of 
the 2-group anomaly to that of non-invertible symmetries.

\section{Summary and discussions\label{sum}}
In this paper, we have studied the non-invertible global 
symmetries and their correlations in $(3+1)$-dimensional 
axion electrodynamics with a massless axion and a massless photon.
We have shown that there is the non-invertible 
1-form symmetry parameterized by a rational number.
This can be formulated by modifying a topological term 
$\phi da $ by using the TQFT 
describing the $(1+1)$-dimensional Schwinger model.

After constructing the non-invertible 1-form symmetry, 
we have discussed the correlation functions between
non-invertible 0- and 1-form symmetry defects.
We have found that the intersections of 0- and 1-form symmetry 
defects induce the invertible magnetic 1- and 2-form symmetry 
generators with boundaries.
We identify the boundaries of the symmetry generators 
as fractionally charged Wilson loops and local axion operators,
and detect their charges by further intersecting 
the 1- and 0-form symmetry generators, respectively.
We relate the correlation functions 
to the Sikivie effect, anomalous Hall effect, and 
axion production on the magnetic flux.
In particular, the fractional charges 
captured by the symmetry transformations 
coincide with those calculated by the equations of motion.

There are several avenues for future work.
We have assumed that the rational number parameterizing 
the non-invertible 1-form symmetry defect is 
given by $1/q_a$ with $q_a \in \bb{Z}$.
It will be possible to extend the parameter to 
general rational number by using a similar approach 
in the case of non-invertible 0-form symmetry
discussed in Ref.~\cite{Choi:2022jqy,Cordova:2022ieu}.
To understand the structure of the non-invertible symmetries, 
it will be important to specify the fusion rule of 
non-invertible 0- and 1-form symmetry defects.

As mentioned in section~\ref{corr},
the fractional charge on the intersection of non-invertible defects may be related to an 't Hooft anomaly.
To see the 't Hooft anomaly of non-invertible symmetries 
systematically, 
it will be important to establish the background gauging 
of the non-invertible symmetries in axion electrodynamics.

\subsection*{Acknowledgments}
The author thanks Kantaro Ohmori for helpful discussions.
This work is supported by JSPS KAKENHI Grants No.~JP21J00480, JP21K13928.

\appendix
\section{Continuous deformation of non-invertible defects\label{top}}

In this appendix, we summarize 
continuous deformations of the non-invertible symmetry defects
and actions on the charged objects.
\subsection{Continuous deformation of 
non-invertible 0-form symmetry defect \label{top0}}
Here, we review a continuous deformation of 
the non-invertible 0-form symmetry defect
$ D_\phi (e^{2 \pi i /(N q_\phi)}, {\cal V})$.
We consider the following correlation function,
\begin{equation}
  \vevs{ D_\phi (e^{2 \pi i /(N q_\phi)}, {\cal V}) }
=
{\cal N} 
\int {\cal D}[\phi, a]e^{iS[\phi, a]}
 e^{i \int_{\cal V} -\fr{2\pi v^2 }{N q_\phi}*d\phi}
\int {\cal D}c \, 
e^{
\fr{i q_\phi}{4\pi} \int_{\cal V} c \wed dc
- \fr{i }{2\pi}
\int_{\cal V} c \wed da 
},
\end{equation}
and 
we continuously deform ${\cal V} $ to ${\cal V}'$.
We introduce a 4-dimensional subspace $\Omega_{{\cal V}, {\cal V}'}$ 
such that 
$\der \Omega_{{\cal V}, {\cal V}'} = {\cal V} \sqcup \b{\cal V}'$.
Under the continuous change, 
we assume that there are no singularities 
by axionic strings or 't Hooft loops where
$a$ or $\phi$ cannot be well-defined.
Since 
$\delta_1 ({\cal V}) = 
\delta_1 ({\cal V}') + d\delta_0 ( \Omega_{{\cal V}, {\cal V}'})$,
the correlation function can be written as
\begin{equation}
\begin{split}
& \vevs{ D_\phi (e^{2 \pi i /(N q_\phi)}, {\cal V}) }
\\
&
=
{\cal N} 
\int {\cal D}[\phi, a]e^{iS[\phi, a]}
 e^{i \int_{\cal V'} -\fr{2\pi v^2 }{N q_\phi}*d\phi}
 e^{-
i \int_{\Omega_{{\cal V}, {\cal V}'}} 
\fr{2\pi v^2 }{N q_\phi}d * d\phi}
\int {\cal D}c \, 
e^{
\fr{i }{4\pi} \int_{\cal V'} (q_\phi c \wed dc -2 c \wed da )
}
\\
&\quad
\times
\int
 {\cal D}c \, 
e^{
\fr{i}{4\pi} \int_{\Omega_{{\cal V}, {\cal V}'}} 
 (q_\phi dc \wed dc - 2dc \wed da )
}, 
\end{split}
\end{equation}
Here, we have taken an extension of $c$ in ${\cal V}$
to $\Omega_{{\cal V}, {\cal V}'}$ and ${\cal V}'$.
This is possible because the term
$e^{\fr{i}{4\pi} \int_{\cal V} 
 (q_\phi c \wed dc - 2c \wed da )}$
does not depend on the choice of an auxiliary 4-dimensional space
whose boundary is ${\cal V}$.
Since there are no singularities in $\Omega_{{\cal V}, {\cal V}'}$,
we can solve the partition function 
$\int
 {\cal D}c \, 
e^{
\fr{i}{4\pi} \int_{\Omega_{{\cal V}, {\cal V}'}} 
 (q_\phi dc \wed dc - 2dc \wed da )
} = e^{
-\fr{i}{4\pi q_\phi} \int_{\Omega_{{\cal V}, {\cal V}'}} 
 da \wed da 
}$.
Finally, by using the redefinition 
$\phi \to 
\phi
+ \fr{2\pi }{N q_\phi } \delta_0 ( \Omega_{{\cal V}, {\cal V}'})$,
we have 
\begin{equation}
\begin{split}
 & 
\vevs{ D_\phi (e^{2 \pi i /(N q_\phi)}, {\cal V}) }
\\
&
=
{\cal N} 
\int {\cal D}[\phi, a]e^{iS[\phi, a]}
 e^{i \int_{\cal V'} -\fr{2\pi v^2 }{N q_\phi}*d\phi}
\int {\cal D}c \, 
e^{
\fr{i }{4\pi} \int_{\cal V'} (q_\phi c \wed dc -2 c \wed da )
}
\\
&
=
\vevs{ D_\phi (e^{2 \pi i /(N q_\phi)}, {\cal V}') }.
\end{split} 
\end{equation}
The correlation function means that 
we can freely move ${\cal V}$ to another subspace ${\cal V}'$
as long as the topology of ${\cal V}$ is not changed.

It is then straightforward to obtain the 
symmetry transformation of the axion operator.
We consider the following correlation function,
\begin{equation}
\begin{split}
 & 
\vevs{ D_\phi (e^{2 \pi i /(N q_\phi)}, {\cal V})e^{i \phi ({\cal P})} }
\\
&
=
{\cal N} 
\int {\cal D}[\phi, a]e^{iS[\phi, a] + i\phi ({\cal P})}
 e^{i \int_{\cal V} 
-\fr{2\pi v^2 }{N q_\phi}*d\phi}
\int {\cal D}c \, 
e^{
\fr{i}{4\pi} \int_{{\cal V}} 
(q_\phi c \wed dc -2 c \wed da )
}\end{split} 
\end{equation}
Under the topological deformation of ${\cal V} $ to ${\cal V}'$,
we redefine the integral variable
$\phi \to 
\phi
+ \fr{2\pi }{N q_\phi } \delta_0 ( \Omega_{{\cal V}, {\cal V}'})$,
and the axion 
operator $e^{i \phi ({\cal P})}$
receives the phase rotation
\begin{equation}
\begin{split}
 & 
\vevs{ D_\phi (e^{2 \pi i /(N q_\phi)}, {\cal V})e^{i \phi ({\cal P})} }
\\
&
=
{\cal N} 
\int {\cal D}[\phi, a]e^{iS[\phi, a] + i\phi ({\cal P})}
 e^{i \int_{{\cal V'} \cup \der\b{\Omega}_{\cal V', V}} 
-\fr{2\pi v^2 }{N q_\phi}*d\phi}
\int {\cal D}c \, 
e^{
\fr{i}{4\pi} \int_{{\cal V'} \cup \der\b{\Omega}_{\cal V', V}} 
(q_\phi c \wed dc -2 c \wed da )
}\\
&
=
{\cal N} 
\int {\cal D}[\phi, a]e^{iS[\phi, a]  + i\phi ({\cal P})}
 e^{i \int_{\cal V'} -\fr{2\pi v^2 }{N q_\phi}*d\phi}
\int {\cal D}c \, 
e^{
\fr{i }{4\pi} \int_{\cal V'} (q_\phi c \wed dc -2 c \wed da )
}
\\
&
=
e^{i  \fr{2\pi }{N q_\phi }  \link ({\cal V} \sqcup \b{\cal V}' ,{\cal P})}
\vevs{ D_\phi (e^{2 \pi i /(N q_\phi)}, {\cal V}') e^{i \phi ({\cal P})}}.
\end{split} 
\end{equation}

\subsection{Continuous deformation of non-invertible 1-form symmetry defect \label{top1}}

Next, we derive a continuous deformation 
of the 1-form symmetry defect
$ D_a  (e^{2 \pi i /(N q_a)}, {\cal S})$. 
We begin with the following correlation function,
\begin{equation}
  \vevs{ D_a  (e^{2 \pi i /(N q_a )}, {\cal S}) }
=
{\cal N} 
\int {\cal D}[\phi, a]e^{iS[\phi, a]}
 e^{i \int_{\cal S} \fr{2\pi }{N q_a e^2} *da }
\int {\cal D} [\chi,u] 
e^{ \fr{i}{2\pi } \int_{\cal S} 
(q_a \chi du - \chi da - u \wed d\phi   ) 
},
\end{equation}
and 
we continuously deform ${\cal S} $ to 
another 2-dimensional closed subspace ${\cal S}'$.
We introduce a 3-dimensional subspace ${\cal V}_{{\cal S}, {\cal S}'}$ 
such that 
$\der {\cal V}_{{\cal S}, {\cal S}'} = {\cal S} \sqcup \b{\cal S}'$.
Under the deformation, 
we again assume that there are no singularities 
by axionic strings or 't Hooft loops.
Since 
$\delta_2 ({\cal S}) = 
\delta_2 ({\cal S}') - d\delta_1 ( {\cal V}_{{\cal S}, {\cal S}'})$,
the correlation function can be written as
\begin{equation}
\begin{split}
& \vevs{ D_a  (e^{2 \pi i /(N q_a )}, {\cal S}) }
\\
&
=
{\cal N} 
\int {\cal D}[\phi, a]e^{iS[\phi, a]}
 e^{i \int_{\cal S'} \fr{2\pi }{N q_a e^2} *da 
+ i \int_{\cal V_{S,S'}} \fr{2\pi }{N q_a e^2} d *da}
\int {\cal D} [\chi,u] 
e^{ \fr{i}{2\pi } \int_{\cal S'} 
(q_a \chi du - \chi da - u \wed d\phi   ) 
}
\\
&
\quad
\times
\int {\cal D} [\chi,u] 
e^{\fr{i}{2\pi } \int_{\cal V_{S,S'}} 
(q_a d\chi \wed du - d\chi\wed  da - du  \wed d\phi   ) 
},
\end{split}
\end{equation}
Here, we have taken an extension of $u$ and $\chi$ in ${\cal S}$
to ${\cal V}_{{\cal S}, {\cal S}'}$ and ${\cal S}'$.
This is possible because the term
$e^{ \fr{i}{2\pi } \int_{\cal S} 
(q_a \chi du - \chi da - u \wed d\phi   ) 
}$
does not depend on the choice of an auxiliary 3-dimensional space
whose boundary is ${\cal S}$.
Since there are no singularities in ${\cal V}_{{\cal S}, {\cal S}'}$,
we can solve the partition function 
$\int {\cal D} [\chi,u] 
e^{\fr{i}{2\pi } \int_{\cal V_{S,S'}} 
(q_a d\chi \wed du - d\chi\wed  da - du  \wed d\phi   ) 
}
 = e^{-
\fr{i}{2\pi q_a } \int_{\cal V_{S,S'}} 
 d\phi \wed da }$.
Finally, by using the redefinition 
$a \to 
a
- \fr{2\pi }{N q_a  } \delta_1 ( {\cal V}_{{\cal S}, {\cal S}'})$,
we have 
\begin{equation}
\begin{split}
 & 
\vevs{ D_a  (e^{2 \pi i /(N q_a )}, {\cal S}) }
\\
&
=
{\cal N} 
\int {\cal D}[\phi, a]e^{iS[\phi, a]}
 e^{i \int_{\cal S'} \fr{2\pi }{N q_a e^2} *da }
\int {\cal D} [\chi,u] 
e^{ \fr{i}{2\pi } \int_{\cal S'} 
(q_a \chi du - \chi da - u \wed d\phi   ) 
}
\\
&
=
\vevs{ D_a  (e^{2 \pi i /(N q_a )}, {\cal S}') }.
\end{split} 
\end{equation}
The correlation function again means that 
we can freely move ${\cal S}$ to another subspace ${\cal S}'$
as long as the topology of ${\cal S}$ is not changed.

We can apply the derivation to the 
symmetry transformation of the Wilson loop
that can be described by the correlation function,
\begin{equation}
\begin{split}
  \vevs{ D_a  (e^{2 \pi i /(N q_a )}, {\cal S}) 
e^{i\int_{\cal C} a}
}
&=
{\cal N} 
\int {\cal D}[\phi, a]e^{iS[\phi, a] +i\int_{\cal C} a }
 e^{i \int_{\cal S} \fr{2\pi }{N q_a e^2} *da }
\\
&
\quad
\times
\int {\cal D} [\chi,u] 
e^{ \fr{i}{2\pi } \int_{\cal S} 
(q_a \chi du - \chi da - u \wed d\phi   ) 
},
\end{split} 
\end{equation}
Under the continuous deformation of ${\cal S} $ to ${\cal S}'$
we redefine the integral variable
$a \to 
a
- \fr{2\pi }{N q_a } \delta_1 ( {\cal V}_{{\cal S}, {\cal S}'})$,
and the Wilson loop
receives the phase rotation
\begin{equation}
\begin{split}
&
  \vevs{ D_a  (e^{2 \pi i /(N q_a )}, {\cal S}) 
e^{i\int_{\cal C} a}
}
\\
&
=
e^{i  \fr{2\pi }{N q_a  }  \link ({\cal S} \sqcup \b{\cal S}' ,{\cal C})}
{\cal N} 
\int {\cal D}[\phi, a]e^{iS[\phi, a] +i\int_{\cal C} a }
 e^{i \int_{\cal S'} \fr{2\pi }{N q_a e^2} *da }
\int {\cal D} [\chi,u] 
e^{ \fr{i}{2\pi } \int_{\cal S'} 
(q_a \chi du - \chi da - u \wed d\phi   ) 
}
\\
&
=
e^{i  \fr{2\pi }{N q_a  }  \link ({\cal S} \sqcup \b{\cal S}' ,{\cal C})}
  \vevs{ D_a  (e^{2 \pi i /(N q_a )}, {\cal S}) 
e^{i\int_{\cal C} a }}.
\end{split} 
\end{equation}

\providecommand{\href}[2]{#2}
\end{document}